\documentstyle[epsf,psfig,aps]{revtex}
\begin{document}
\newcommand{\h}{\frac{1}{2}}
\newcommand{\Le}{\left(}
\newcommand{\Ra}{\right)}
\newcommand{\beq}{\begin{equation}}
\newcommand{\dlq}{\lq\lq}
\newcommand{\eeq}{\end{equation}}
\newcommand{\ben}{\begin{eqnarray}}
\newcommand{\een}{\end{eqnarray}}
\newcommand{\stackeven}[2]{{{}_{\displaystyle{#1}}\atop\displaystyle{#2}}}
\newcommand{\lsim}{\stackeven{<}{\sim}}
\newcommand{\gsim}{\stackeven{>}{\sim}}
\renewcommand{\baselinestretch}{1.0}
\newcommand{\as}{\alpha_s}
\newcommand{\NC}{\alpha_{s}^{2}\,\frac{\,N_{c}^{2}\,-\,1\,}{\,N_{c}^{2}\,}}
\def\eq#1{{Eq.~(\ref{#1})}}
\def\fig#1{{Fig.~\ref{#1}}}
\vspace*{1cm} 
\setcounter{footnote}{1}

%%%%%%%%%%%%%%%%%%%%%%%%%%%%%%%%%%%%%%%%%%%%%%%%%%%%%%%%%%%%%%%%%

% ABBREVIATED JOURNAL NAMES  
%
\def\ap#1#2#3{     {\it Ann. Phys. (NY) }{\bf #1} (19#2) #3}
\def\arnps#1#2#3{  {\it Ann. Rev. Nucl. Part. Sci. }{\bf #1} (19#2) #3}
\def\npb#1#2#3{    {\it Nucl. Phys. }{\bf B#1} (19#2) #3}
\def\plb#1#2#3{    {\it Phys. Lett. }{\bf B#1} (19#2) #3}
\def\prd#1#2#3{    {\it Phys. Rev. }{\bf D#1} (19#2) #3}
\def\prep#1#2#3{   {\it Phys. Rep. }{\bf #1} (19#2) #3}
\def\prl#1#2#3{    {\it Phys. Rev. Lett. }{\bf #1} (19#2) #3}
\def\ptp#1#2#3{    {\it Prog. Theor. Phys. }{\bf #1} (19#2) #3}
\def\rmp#1#2#3{    {\it Rev. Mod. Phys. }{\bf #1} (19#2) #3}
 \def\zpc#1#2#3{    {\it Z. Phys. }{\bf C#1} (19#2) #3}
\def\mpla#1#2#3{   {\it Mod. Phys. Lett. }{\bf A#1} (19#2) #3}
\def\nc#1#2#3{     {\it Nuovo Cim. }{\bf #1} (19#2) #3}
\def\yf#1#2#3{     {\it Yad. Fiz. }{\bf #1} (19#2) #3}
\def\sjnp#1#2#3{   {\it Sov. J. Nucl. Phys. }{\bf #1} (19#2) #3}
\def\jetp#1#2#3{   {\it Sov. Phys. }{JETP }{\bf #1} (19#2) #3}
\def\jetpl#1#2#3{  {\it JETP Lett. }{\bf #1} (19#2) #3}
%%%%%%%%% notice the parenthesys is only on one side
\def\ppsjnp#1#2#3{ {\it (Sov. J. Nucl. Phys. }{\bf #1} (19#2) #3}
\def\ppjetp#1#2#3{ {\it (Sov. Phys. JETP }{\bf #1} (19#2) #3}
\def\ppjetpl#1#2#3{{\it (JETP Lett. }{\bf #1} (19#2) #3} 
\def\zetf#1#2#3{   {\it Zh. ETF }{\bf #1}(19#2) #3}
\def\cmp#1#2#3{    {\it Comm. Math. Phys. }{\bf #1} (19#2) #3}
\def\cpc#1#2#3{    {\it Comp. Phys. Commun. }{\bf #1} (19#2) #3}
\def\dis#1#2{      {\it Dissertation, }{\sf #1 } 19#2}
\def\dip#1#2#3{    {\it Diplomarbeit, }{\sf #1 #2} 19#3 }
\def\ib#1#2#3{     {\it ibid. }{\bf #1} (19#2) #3}
\def\jpg#1#2#3{        {\it J. Phys}. {\bf G#1}#2#3}  

\begin{center}
{\Large\bf `Effective' Pomeron model at large $\mathbf{b}$}
\\[1cm]
Sergey Bondarenko \,\\ ~~ \\
{\it  HEP Department, School of Physics and Astronomy } \\
{\it Tel Aviv University, Tel Aviv 69978, Israel } \\ ~~ \\
\end{center}

\begin{abstract}
In this paper we consider 
the influence of non-perturbative
corrections on the large $\,b\,$ (impact parameter)
behavior of the BFKL amplitude.
This is done in the framework of a 
model where such ``soft'' corrections are taken into account
in the BFKL kernel.
We show, that
these corrections lead to a power-like decreasing behavior
of amplitude, which differs from the BFKL case.
\end{abstract}

\subsection{Introduction}

In this paper we consider the influence 
of the non-perturbative corrections to the behavior of the perturbative
amplitude at large $\,b\,$ (impact parameter).
We do not have a consistent theoretical approach
for taking into account non-perturbative
corrections  in QCD, so we examine our problem 
in the framework of an `effective' Pomeron model
introduced in \cite{KLK}, \cite{BON}. In this model we take
for the  perturbative amplitude 
the solution of the BFKL equation in leading order of QCD
with fixed coupling constant \cite{B_BFKL}. Here the `effective'
Pomeron appears as the solution
of the Bethe-Salpeter equation, resulting from the ``ladder'',
where a `soft' kernel is included,
together with  the BFKL kernel. 
Therefore, this Pomeron represents a  particular
example of including non-perturbative corrections in the QCD
high energy scattering  amplitude.

It is well known, that the  BFKL amplitude has a power-like decreasing
behavior at large $\,b\,$, see \cite{LIP}, \cite{NAV}, \cite{K-L-B}. 
Such behavior contradicts the general postulates
of analyticity and crossing symmetry. From these postulates
we know, that the
absence in the hadron spectrum of particles with zero mass, dictates
exponential $\,e^{-2\,b\,m_{\pi}}\,$ decrease of the scattering amplitude,
see \cite{Unit}. At the same time, the `soft' Pomeron, 
as a Regge trajectory, provides sharp exponential
$\,e^{-b^2\,B\,}\,$ decreasing behavior of the amplitude. 
There are different approaches which were suggested 
so as  to achieve such an exponential decrease,
see \cite{K-L-B}, \cite{JEN}, \cite{KOV}, \cite{HET}.
In papers \cite{K-L-B}, \cite{JEN} it was suggested, that
to obtain the necessary result
it is sufficient to include the non-perturbative corrections only
in the Born term  of the BFKL amplitude, 
leaving the kernel unchanged.
On the other hand, there is a point of view, which claims that
such a decrease may be achieved by only  including  
non-perturbative QCD corrections 
in the BFKL kernel, see \cite{KOV}. 
The model, considered in this paper, 
gives a particular example of such an approach, but 
will not lead to the
unitarization of the cross section.
Indeed, adding short range correction we do not modify
the large $\,b\,$ behavior of the BFKL kernel, that
necessary for the unitarization of the amplitude.
But exploring the large $\,b\,$ behavior of the amplitude,
which is the 
admixture of  the non-perturbative and the  BFKL kernels,
we  investigate  the differences between this 
amplitude and the BFKL one at large $\,b\,$,
that clarify the importance of the non-perturbative corrections
even in the usual high-energy processes of pQCD.

  In the next section we consider the 
BFKL amplitude for  $\,q\,\neq\,0\,$.
In the following sections this amplitude will be
used to construct the `effective' Pomeron,
and only the diffusion approximation
for this `hard' amplitude will be considered.
In section 3, we consider the `effective' Pomeron in the 
$\,q\,$ representation.
In section 4 we discuss the $\,b\,$ representation of
the amplitude which was found in the section 3.
In section 5 we compare the  large $\,b\,$
behavior of the BFKL and `effective' Pomeron.
In the last section we summarize our results.

\subsection{Dipole amplitude in the diffusion approach}

We start this section considering  the dipole-dipole 
BFKL amplitude, which for large $\,b\,$ has the form:

\beq\label{B-B} 
N^{BFKL}(y,r_{1, t},r_{2, t}; b)\,\,=
\,\,\int \,\frac{d \nu}{2\,\pi}\,
\phi_{in}\,
e^{\omega(\nu)\,y} \,2^{4\,-\,8\,i\,\nu}\,
\,\left(\,\frac{r^2_{1,t}\,r^2_{2,t}}
{\,b^4\,}\,\right)^{\h + i\,\nu}\,\,,
\eeq
where  $\,b\,>\,r_{2,t}\,>\,r_{1,t}\,$, see \cite{LIP}, \cite{NAV}.
The factor $\,2^{4\,-\,8\,i\,\nu}\,\,$ is the
scale factor which is determined by the conformal 
invariance of the solution of the  BFKL equation, see also
\cite{P-M}, \cite{Sal}. The \eq{B-B} is divergent at small 
values of $\,b\,$ due the fact that
we use the approximate solution at large $\,b\,$ for the
BFKL  Green function.
The correct solution of the BFKL equation, of course, 
gives finite result for all range of $\,b\,$,  
but has not been used  in the treatment of the problem due the 
mathematical difficulties of such consideration.     
To avoid this divergence arising in integration over $\,b\,$
in our calculations, we  will be use the following 
expression instead of \eq{B-B}:

\beq \label{Ampl1}
N(y,r_{1, t},r_{2, t}; b)\,\,=
\,\,\int \,\frac{d \nu}{2\,\pi}\,
\phi_{in}\,
e^{\omega(\nu)\,y} \,2^{4\,-\,8\,i\,\nu}\,
\,\left(\,\frac{r^2_{1,t}\,r^2_{2,t}}
{(\,b^2\,+\,r_{2,t}^{2}/4\,)^2\,}\,
\right)^{\h + i\,\nu}\,\,.
\eeq 
We see that the  BFKL amplitude \eq{B-B} has approximately a 
$\,1/b^4\,$ behavior  at large $\,b\,$, 
and in  integration over $\,b\,$ such behavior leads to the  divergence
of the integral in the region of small $\,b\,$.
Therefore, we introduce a cut over $\,b\,$ in \eq{Ampl1},
writing $\,(b^2\,+\,r_{2}^{2}/4)\,$ instead
$\,b^2\,$ in denominator of expression. This substitution
has no  affect on the results obtained at
large $\,b\,$, see \cite{K-L-B}.
The form of $\phi_{in}(r_{1,t}; b)$ is defined by the
large $\,b\,$ behavior of the Born term of the BFKL amplitude:

\beq \label{Ampl2}
\phi_{in}\,=\,\pi\,\NC\,\frac{1}{1/2\,-\,i\,\nu\,}\,.
\eeq
In \eq{Ampl1}   $\,\omega(\nu)\,$ is the 
eigenvalue of the full BFKL kernel. 
In the following, to simplify the
calculations, 
we will use the  diffusion approximation for $\,\omega(\nu)\,$, 
( where
$\,\,\omega(\nu)\,=\,\omega_0\,-D\,\nu^2\,$),
which is valid at small values of $\,\nu\,$.
Together \eq{Ampl1} and \eq{Ampl2} give:

\beq \label{Ampl3}
N(y,r_{1, t},r_{2, t}; b)\,\,=
\,\pi\,\NC\,\int \,d \nu\,\frac{e^{\omega(\nu)\,y}\,
\,2^{4\,-\,8\,i\,\nu}\,}{2\,\pi\,(1/2\,-\,i\,\nu\,)}
\,\left(\,\frac{r^2_{1,t}\,r^2_{2,t}}{(\,b^2\,+\,r_{2,t}^{2}/4\,)^2\,}\,
\right)^{\h + i\,\nu}\,.
\eeq
The  Born term of \eq{Ampl3} with the $\,\phi_{in}\,$ of \eq{Ampl2} is :

\beq \label{Ampl33}
N(y\,=\,0\,,r_{1, t},r_{2, t}; b)_{B}\,\,=\,
\pi\,\NC\,\frac{r^2_{1,t}\,r^2_{2,t}}{(\,b^2\,+\,r_{2,t}^{2}/4\,)^2\,}\,\,,
\eeq
see \cite{K-L-B}.

The `effective' Pomeron `ladder' , see Fig.~\ref{S-H}, was studied
in Ref.\cite{BON} at $\,q\,=\,0\,$, the generalization to  $\,q\,\neq\,0\,$,
which we need here, is very simple. Indeed, the equation
that sums `ladder' diagrams is 
of Bethe-Salpeter type, and  the momentum transferred
$\,q\,$, is a parameter which is preserved along a `ladder'. 
Therefore, the equation 
for $\,q\,\neq\,0\,$  can be written in the same way as for 
$\,q\,=\,0\,$, by introducing the kernel with $\,q\,$
dependence. Doing so, the form of solution will be the same
as for $\,q\,=\,0\,$, see next section. Consequently, in the further 
calculations we need to know the BFKL amplitude written in the $\,q\,$
representation:

\beq \label{Ampl4}
N(y,r_{1, t},r_{2, t}; q)\,\,=\,
\eeq
$$
\pi\,\NC\,
\int \,d \nu\,\frac{e^{\omega(\nu)\,y}\,
\Le\,r^2_{1,t}\,r^2_{2,t}\,\Ra^{\h + i\,\nu}\,
\,2^{4\,-\,8\,i\,\nu}\,}
{2\,\pi\,(1/2\,-\,i\,\nu\,)}\,
\int\,\frac{d\,b}{2\,\pi}\,J_{0}(|q|\,|b|)\,
\frac{b}{(\,b^2\,+\,r_{2,t}^{2}/4\,)^{1\,+\,2\,i\,\nu}}\,=
$$
$$
\,=\,\NC\,
\Le\,\,r_{1,t}\,r_{2,t}\,\Ra\,\int\,
d\,\nu\,\frac{e^{\omega(\nu)\,y}\,
\Le\,r^2_{1,t}\,q^2\,\Ra^{\,i\,\nu}\,\,2^{4\,-\,8\,i\,\nu}\,}
{2\,\pi\,(1/2\,-\,i\,\nu\,)}\,
\frac{K_{2\,i\,\nu}\,(|q|\,|r_{2,t}|/2)}{\Gamma(1\,+\,2\,i\,\nu)\,}\,\,,
$$ 
where $\,K_{2\,i\,\nu}\,$ is the McDonald function.
In the limit of the Born term at $\,y\,=\,0\,$,
performing contour integration and returning
to the $\,b\,$ representation, 
we again obtain the expression given by \eq{Ampl33}.
Finally we have:

\beq \label{Ampl5}
N(y,r_{1, t},r_{2, t}; q)\,\,=
\int\,\frac{d\,\omega}{2\,\pi\,i}\,\,
e^{\omega\,y\,}\,N^{\omega}(r_{1, t},r_{2, t}; q)\,\,=\,
\eeq
$$
\,\NC\,\Le\,\,r_{1,t}\,r_{2,t}\,\Ra\,
\int\,\frac{d\,\omega}{2\,\pi\,i}\,\int\,
\frac{d\,\nu\,}{2\,\pi}\frac{e^{\omega\,y}\,
\Le\,r^2_{1,t}\,q^2\,\Ra^{\,i\,\nu}\,2^{4\,-\,8\,i\,\nu}\,}
{(\omega\,-\,\omega(\nu))\,\,(1/2\,-\,i\,\nu\,)}\,
\frac{K_{2\,i\,\nu}\,(|q|\,|r_{2,t}|/2)}{\Gamma(1\,+\,2\,i\,\nu)\,}\,.
$$
In the  further  calculations we will use 
$\,N^{\omega}(r_{1, t},r_{2, t}; q)\,$ given by \eq{Ampl5}.

\subsection{`Effective' Pomeron}

\begin{figure}
\begin{center}
\epsfxsize=5cm
\leavevmode
\hbox{ \epsffile{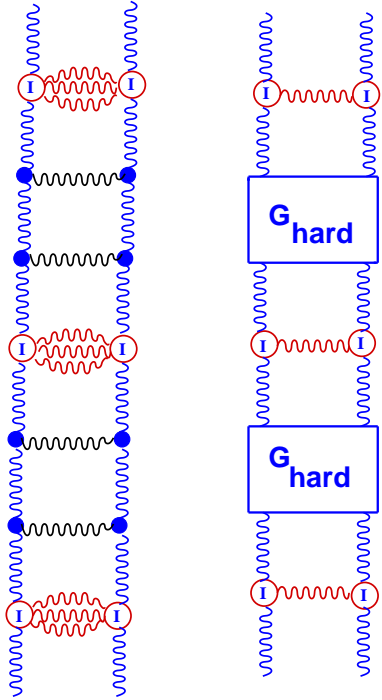}}
\end{center}
\caption{The diagrams and the graphic form of the `admixture'
of `soft' and `hard' kernels.}
\label{S-H}
\end{figure}

In this section  we consider the approach for the  `effective' Pomeron, 
which was introduced in  the Refs.\cite{KLK},\cite{BON}.
The main idea behind this Pomeron
is shown in Fig.~\ref{S-H}.
The resulting solution for the `ladder' of Fig.~\ref{S-H},
can be found as a solution of a  Bethe-Salpeter type
equation, and this solution is

\beq\label{GEN}
N_{full}^{\omega}\,(r_{1, t},r_{2, t}; q)\,=\,
N^{\omega}\,(r_{1, t},r_{2, t}; q)\,+\,
\frac{\,\tilde{N}^{\omega}(r_{1, t},r_{2, t}; q)\,}
{1\,-\,A\int\,d^{2}\,k_1\,\int\,d^{2}\,k_2\,
\phi(k_1,q)\,k_1^2\,
N_{0}^{\omega}(k_{1},k_{2}; q)\,k_2^2\,\phi(k_2,q)\,}\,,
\eeq 
where is $\,\tilde{N}^{\omega}(r_{1, t},r_{2, t}; q)\,$
is defined in Appendix C, and where $\,N_{0}^{\omega}(k_{1},k_{2}; q)\,$ in
\eq{SH1} is the Fourier transform of 
\beq\label{New}
N_{0}^{\omega}(r_{1, t},r_{2, t}; q)\,
=\,
\,\Le\,\,r_{1,t}\,r_{2,t}\,\Ra\,
\int\,
\frac{d\,\nu\,}{2\,\pi}\frac{\,
\Le\,r^2_{1,t}\,q^2\,\Ra^{\,i\,\nu}\,2^{4\,-\,8\,i\,\nu}\,}
{(\omega\,-\,\omega(\nu))\,\,(1/2\,-\,i\,\nu\,)}\,
\frac{K_{2\,i\,\nu}\,(|q|\,|r_{2,t}|/2)}{\Gamma(1\,+\,2\,i\,\nu)\,}\,\,.
\eeq
The second term of \eq{GEN} is the
`effective' Pomeron, which is an  
admixture of `soft'
and `hard' kernels written in the $\,q\,$ representation:

\beq\label{SH1}
N_{S-H}^{\omega}\,(r_{1, t},r_{2, t}; q)\,=\,
\frac{\,\tilde{N}^{\omega}(r_{1, t},r_{2, t}; q)\,}
{1\,-\,A\int\,d^{2}\,k_1\,\int\,d^{2}\,k_2\,
\phi(k_1,q)\,k_1^2\,
N_{0}^{\omega}(k_{1},k_{2}; q)\,k_2^2\,\phi(k_2,q)\,}\,,
\eeq
Throughout the rest of  the paper we will concentrate on
this part of the full solution for the `effective ladder'.
\eq{SH1} has a different form than the expression
obtained in \cite{BON}.
Firstly, as mentioned in the previous section,
the function of \eq{SH1}
depends on $\,q\,$, whereas in \cite{BON} it
was obtained as a  solution for the case $\,q\,=\,0\,$. 
Another  difference between \eq{SH1} and the solution in 
paper \cite{BON}, is in the form of the `hard' Green's function. 
The rank of our `effective 
ladder' in  Fig.~\ref{S-H} contains two propagators,
one from the soft kernel and another from the hard, 
see Fig.~\ref{S-H}. In \cite{BON}
the fully truncated Green's function was used.
Therefore, in Ref.\cite{BON}, the additional propagator
$\,1/k^2\,$ was included in integrations over each internal
momenta.
In our case, the function of \eq{Ampl1} 
has four external propagators, and ,therefore, we have to truncate
two of them. We do so, including $\,k^2\,$ in the integration
over this internal momenta.

We also use
the soft kernel of the model 
$\,K_{soft}(k_1\,,k_2\,,q)\,=\,\Delta_S\,\phi(k_1,q)\,\phi(k_2,q)\,$,
where
\beq\label{SH2}
\phi(k,q)\,=\,e^{\,-\frac{k^2}{2\,q^{2}_{s}}\,-
\,\frac{(\vec{q}\,-\,\vec{k})^2}{2\,q^{2}_{s}}}\,\,.
\eeq
In the following calculations, for the sake of simplicity,
we take the kernel in this  form,
but this form of the kernel is also dictated by the instanton approach,
see \cite{KLK}, and could have a more general basis.
We will find later the numerical factor
$\,A\,$ in \eq{SH1}, which
provides the correct soft pole position in denominator of \eq{SH1}.
Performing  the Fourier transform
for the transverse momenta $\,k_1\,$, $\,k_2\,$   to the 
size of dipoles $\,r_{1, t}\,$,  $\,r_{2, t}\,$
in the denominator of \eq{SH1}, 
we obtain:

\beq\label{SH11}
N_{S-H}^{\omega}\,(y,r_{1, t},r_{2, t}; q)\,=\,
\eeq
$$
\,=\,\frac{\,\tilde{N}^{\omega}(y,r_{1, t},r_{2, t}; q)\,}
{1\,-\,A\int\,d^{2}\,k_1\,k_1^2\,\int\,d^{2}\,k_2\,k_2^2\,
\phi(k_1,q)\,\,\phi(k_2,q)\,
\int\,\frac{d^{2}\,r_{1,t}\,}{(2\,\pi)^2}\,
e^{-i\,\vec{k}_1\,\vec{r}_1}\,
\int\,\frac{d^{2}\,r_{2,t}\,}{(2\,\pi)^2}\,
e^{-i\,\vec{k}_2\,\vec{r}_2}\,
N_{0}^{\omega}(r_{1, t},r_{2, t}; q)\,}\,.
$$

We find the constant $\,A\,$ of \eq{SH1} considering  the  
Born term for  the BFKL
amplitude in $\,q\,$ representation:

$$
N_{0}^{\omega}(k_{1},k_{2}; q)\,=\,
\frac{\delta^{2}(\,\vec{k}_2\,-\,\vec{q}\,+\,\vec{k}_1\,)}
{k_1^2\,k_2^2}\,.
$$
We obtain:

\beq\label{SH4}
\frac{A}{\omega}\,\int\,d^{2}\,k_1\,\int\,d^{2}\,k_2\,
\phi(k_1,q)\,\delta^{2}(\,\vec{k}_2\,-\,\vec{q}\,+\,\vec{k}_1\,)\,
\phi(k_2,q)\,=\,
\eeq
\beq\label{SH5}
\frac{A}{\omega}\,\int\,d^{2}\,k_1\,
e^{\,-\frac{k^{2}_{1}}{q^{2}_{s}}\,-
\,\frac{(\vec{q}\,-\,\vec{k}_1)^2}{q^{2}_{s}}}\,=\,
\frac{A}{2\,\omega}\,\pi\,q^{2}_{s}\,e^{-q^2\,/\,2\,q^{2}_{s}\,}\,.
\eeq
From Ref.\cite{BON} we know,
that in this case, the `effective' amplitude  has the form:

\beq\label{SH55}
\frac{1}{\omega}\frac{\,K_S\,/\,q_s^2}
{\,\omega\,-\,\Delta_{S}\,+\,\alpha^{'}\,q^2\,\,}.
\eeq
We consider here   the case of small $\,q\,$,
$\,\frac{\,q^2\,}{\,q^{2}_{s}\,}\,<\,1\,\,$.
The  comparison with \eq{SH5} gives:

\beq\label{SH555}
\frac{A}{2}\,\pi\,q^{2}_{s}\,e^{-q^2\,/\,2\,q^{2}_{s}\,}\,=\,
\,\Delta_{S}\,-\,\alpha^{'}\,q^2\,\,,
\eeq 
and
\beq\label{SH6}
A\,=\,\frac{2\,\Delta_{S}}{\pi\,q^{2}_{s}}\,,
\,\,
\alpha^{'}\,=\,\frac{A\,\pi}{4}\,=\,\frac{\Delta_{S}}{2\,q^{2}_{s}}\,.
\eeq
These expressions show the relationships
of the parameters of the `soft' Pomeron
trajectory with the `soft' kernel given
by \eq{SH2}.

The `effective' pole of \eq{SH1}
is the solution of the following equation:

\beq\label{Pole}
1-\frac{2\,\Delta_{S}}{\pi\,q^{2}_{s}}
\int\,d^{2}k_1\,k_1^2\int\,d^{2}k_2\,k_2^2\,
\phi(k_1,q)\,\phi(k_2,q)\,
\int\frac{d^{2}\,r_{1,t}}{(2\pi)^2}
e^{-i\vec{k}_1\vec{r}_1}
\int\frac{d^{2}\,r_{2,t}}{(2\pi)^2}
e^{-i\vec{k}_2\vec{r}_2}
N_{0}^{\omega}(r_{1, t},r_{2, t}; q)=0\,.
\eeq
This equation is solved 
in Appendix A, and the solution obtained is:

\begin{enumerate}
\item In the region where
$\,2\,\nu_0\,<\,1\,$
and $\,2\,\nu_0\,\ln(\frac{q_s^2}{q^2})\,<\,1\,$
with

\beq\label{Nu1}
\nu_0\,=\,\frac{4\,\Delta_S}{D}\,
\ln(\frac{q_s^2}{q^2})\,\,,
\eeq
the solution of \eq{Pole} is given by

\beq\label{HS2}
\omega\,=\,\omega_{S-H}\,=\,\omega_0\,+\,
\Le\,\Delta_S\,\Ra^{2}\,
\frac{\ln^{2}(\frac{q_s^2}{q^2})\,}{\,2^{\,16\,\nu_0\,-\,4}\,D}\,.
\eeq

\item In the region
$\,2\,\nu_0\,<\,1\,$
and $\,2\,\nu_0\,\ln(\frac{q_s^2}{q^2})\,>\,1\,$,
where

\beq\label{Nu2}
\nu_0\,=\,
\sqrt{\,\frac{2\,\Delta_S}{D}\,}\,,
\eeq

the solution for `effective' pole reads:

\beq\label{HS3}
\omega\,=\,\omega_{S-H}\,=\,\omega_0\,+\, 
\,\frac{\Delta_S}{\,2^{\,8\,\nu_0\,-\,1}\,}\,-\,
\,2\,\Delta_S\,\Le\,\frac{q^2}{16\,q_s^2}\,\Ra^{2\nu_0}\,.
\eeq

\end{enumerate}
Below we denote both solutions by $\,\omega_{S-H}\,$,
unless mentioned otherwise.

Returning  to the \eq{SH1}, 
and integrating this equation over $\,\omega\,$
we obtain:

\beq\label{SH20}
N_{S-H}(y,r_{1, t},r_{2, t}; q)\,=\,
\int\,\,\frac{d\,\omega}{2\,\pi\,i}\,e^{y\,\omega}\,
N_{S-H}^{\omega}\,(r_{1, t},r_{2, t}; q)\,=\,
\eeq
$$
\int\,\,\frac{d\,\omega}{2\,\pi\,i}\,
\frac{\,f(\omega_{S-H}\,,\omega\,)}{\omega\,-\,\omega_{S-H}}
\,e^{\omega\,y}\,
\tilde{N}^{\omega}(r_{1, t},r_{2, t}; q)\,,
$$
with the 
$\,\tilde{N}^{\omega}(r_{1, t},r_{2, t}; q)\,$
defined by \eq{C4} and 
$\,f(\omega_{S-H}\,,\omega\,) = 
2\sqrt{\omega-\omega_0}\sqrt{\omega_{S-H}-\omega_0}\, $ 
in the case of solution \eq{HS2}, or 
$\,f(\omega_{S-H}\,,\omega\,) = 
\omega-\omega_0\,$  in the case of solution \eq{HS3}.

For high energies we have that 
$\,\omega_{S-H}\,-\,\omega_0\,>\,D\,\nu_{SP}^{2}\,$ ,
where  $\,\nu_{SP}\,=\,\frac{\ln(q^2/q_s^2)}{2\,D\,y}\,$
or  $\,\nu_{SP}\,=\,\frac{\ln(r_{1,t}q_s)}{2\,D\,y}\,$,
are the values of
saddle points in integration over $\,\nu_1\,$ or $\,\nu_2\,$  in  \eq{C4}
by the method of steepest descent.
Therefore, after the integration over $\,\omega\,$, we have :

\beq\label{HS4}
N_{S-H}(y,r_{1, t},r_{2, t}; q)\,=\,
\,f(\omega_{S-H}\,,\omega_{S-H}\,)\,
\,e^{\omega_{S-H}\,y}\,
\tilde{N}^{(\omega=\omega_{S-H})}(r_{1, t},r_{2, t}; q)\,.
\eeq

We can calculate 
$\,\tilde{N}^{(\omega=\omega_{S-H})}(r_{1, t},r_{2, t}; q)\,\,$,
see Appendix D. In approximation of small $\,\nu\,$ 
and $\,\frac{q}{q_s}\,<\,1\,$
we obtain

\beq\label{HS5}
N_{S-H}(y,r_{1, t},r_{2, t}; q)\,=\,
C\,\pi\,\Delta_S\,\,e^{\omega_{S-H}\,y}\,
\frac{\Le\,\,r_{1,t}\,r_{2,t}\Ra}{16^{4\,\nu_0\,-\,1}\,D\,\nu_0}\,
\Le\,\frac{q}{q_s}\,\Ra^{2\,\nu_0}\,
\Le\,r_{1,t}\,q_s\,\Ra^{2\,\nu_0}\,
K_{2\,\nu_0}\,(|q|\,|r_{2,t}|/2)\,.
\eeq
where 
$\,\nu_0\,=\,\sqrt{\frac{\omega_{S-H}-\omega_0}{D}}\,$ and
$\,C=2\,$ in the case of solution \eq{HS2}, or $\,C=\,
1\,$ in the case of solution \eq{HS3}.
We see, that instead of the saddle point value of $\,\nu\,$
in \eq{Ampl4}, the small $\,q\,$ behavior of \eq{HS5}
is governed by non-perturbative contributions in $\,\nu\,$,
which are given by \eq{Nu1} or by  \eq{Nu2}.

\subsection{Large $\mathbf{b}$ behavior of `effective' Pomeron}

In this section we will consider the large $\,b\,$ 
representation of the amplitude \eq{HS5}.
We define $\,N_{S-H}(y,r_{1, t},r_{2, t}; b)\,$
as

\beq\label{B1}
\,N_{S-H}(y,r_{1, t},r_{2, t}; b)\,=
\int^{q_s}_{0}\,d^2\,q\,e^{i\,\vec{q}\,\vec{b}\,}
N_{S-H}(y,r_{1, t},r_{2, t}; q)\,.
\eeq

The calculation of $\,N_{S-H}(y,r_{1, t},r_{2, t}; b)\,$
is performed in Appendix B.
The results, obtained there, are the following.

\begin{enumerate}
\item In the region of $\,b\,$ where

\beq\label{B4}
4\,b\,q_s\,\Le\,\frac{1}{2\,\Delta_S\,y}\,\Ra^{1/4\nu_0}\,<\,1\,,
\eeq
or
\beq\label{B5}
b\,<\,b_{max}\,=\,\frac{1}{4\,q_s}\,
\Le\,2\,\Delta_S\,y\,\Ra^{1/4\nu_0}\,\,,
\eeq
the amplitude is

\beq\label{B6}
N_{S-H}(y,r_{1, t},r_{2, t}; b)=
\frac{2^{4\nu_0}\pi^2}{16^{4\,\nu_0\,-\,1}}\,\sqrt{\frac{2\Delta_S}{D}}\,
\Le\,\frac{8}{\Delta_S\,y}\,\Ra^{1/2\nu_0}\,
\Le\,r_{1,t}\,r_{2,t}\,q_s^2\Ra\,
\Le\frac{r_{1,t}}{r_{2,t}}\Ra^{2\,\nu_0}\,
e^{\omega_{0}\,y+y\frac{\Delta_S}{2^{8\,\nu_0\,-\,1}}}\,
\Gamma(1+\frac{1}{2\nu_0})\,,
\eeq
with 
\beq\label{B7}
\nu_0\,=\,\sqrt{\frac{2\,\Delta_S}{D}}\,.
\eeq
We see, that for such $\,b\,$, the amplitude does not depend on
$\,b\,$.

\item For impact parameters  which are such that

\beq\label{B8}
b\,>\,b_{max}\,=\,\frac{1}{4\,q_s}\,
\Le\,2\,\Delta_S\,y\,\Ra^{1/4\nu_0}\,\,,
\eeq
the amplitude reads as

\beq\label{B9}
N_{S-H}(y,r_{1, t},r_{2, t}; b)=
\frac{2\pi^2}{16^{4\,\nu_0\,-\,1}}\,\sqrt{\frac{8\Delta_S}{D}}\,
\frac{\Le\,r_{1,t}\,r_{2,t}\,\Ra^{1+2\nu_0}\,}
{\Le\,b^2\,+\,r_{2,t}^2/4\,\Ra^{1+2\nu_0}\,}\,
e^{\omega_{0}\,y+y\frac{\Delta_S}{2^{8\,\nu_0\,-\,1}}}\,
\Gamma(1+2\nu_0)\,,
\eeq
with the same $\,\nu_0\,$ as in \eq{B7}.

\end{enumerate}

\subsection{Large $\mathbf{b}$  behavior of BFKL amplitude}

We return to \eq{Ampl3} and consider the
large $\,b\,$ behavior of this amplitude. The integration over
$\,\nu\,$, may be performed by the method of steepest descent, and 
for large $\,b\,$ the 
answer is:

\beq\label{L1}
N_{BFKL}(y,r_{1, t},r_{2, t}; b)\,\propto\,
\frac{1}{\sqrt{y}}
\frac{r_{1, t}\,r_{2, t}}{b^2\,+\,r_{2, t}^{2}/4\,}\,
e^{\omega_0\,y}\,. 
\eeq

The full expression for the `effective ladder' of Fig.~\ref{S-H},
is given by the sum of the usual
BFKL amplitude and `effective' Pomeron, which are \eq{Ampl3},
\eq{B6} and \eq{B9} correspondingly, see \eq{GEN}. 
From \eq{GEN} it is also clear, 
that when $\,\Delta_S\,=\,0$ we stay only with the BFKL amplitude,
which is  the lower bound for the `effective' Pomeron.  
So, from the value of $\,b\,$ larger than some $\,b_0\,$,
the large $\,b\,$ behavior of the `ladder'
is governed by the BFKL amplitude. The `effective'
Pomeron has a larger intercept than the  BFKL one,
but the BFKL amplitude decreases more slowly at large $\,b\,$. 
We find the value of $\,b_0\,$ by
comparison \eq{B9} and \eq{L1}. It is approximately:

\beq\label{L2}
\frac{1}{\sqrt{y}}\,\simeq\,\Le\,
\frac{r_{1, t}\,r_{2, t}\,}{b_0^2}\,\Ra^{2\,\nu_0}
e^{y\,\frac{\Delta_S}{2^{8\,\nu_0\,-\,1}}}\,,
\eeq
which gives

\beq\label{L3}
b_0\,=\,y^{1/8\nu_0}\,\sqrt{r_{1, t}\,r_{2, t}\,}\,
e^{y\,\frac{\Delta_S}{2^{8\,\nu_0\,+\,1}\,\nu_0}}\,.
\eeq
For small $\,\nu_0\,$ and large $\,y\,$, $\,b_0\,$ 
is a large number, from which the large $\,b\,$
behavior of the full solution for the `effective
ladder' is governed by the BFKL amplitude.

\subsection{Conclusion}

  In this paper we studied the 
influence of non-perturbative corrections on perturbative amplitude
at large  values of  impact parameter. 
We investigated the  influence in the framework of
the model of the `effective' Pomeron, which is   
given by \eq{SH1}.
This Pomeron is the  `admixture' of the pQCD
BFKL amplitude
and the soft, non-perturbative Pomeron, and gives an  example of the model
where non-perturbative corrections are included in the perturbative 
QCD kernel.

\begin{figure}
\begin{center}
\epsfxsize=16cm
\leavevmode
\hbox{ \epsffile{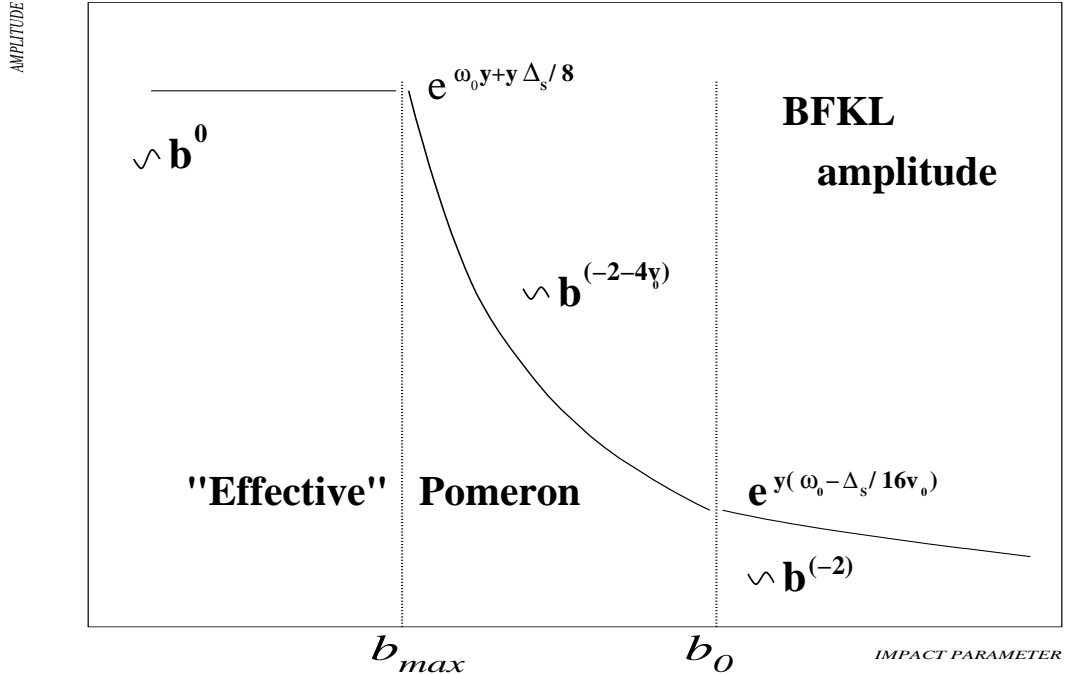}}
\end{center}
\caption{Amplitude behavior in impact parameter space.}
\label{REG}
\end{figure}

The main  result of this paper is given  by
\eq{B6} and \eq{B9}. The impact parameter dependence
of the amplitude,  at large values of impact parameter is the following. 
In the region of $\,b\,$ where

$$
b\,<\,b_{max}\,=\,\frac{1}{4\,q_s}\,
\Le\,2\,\Delta_S\,y\,\Ra^{1/4\nu_0}\,\,,
$$
the $\,b\,$ dependence of amplitude is defined by
\eq{B6}.
For a constant value of rapidity this amplitude is constant,
and does not depend on $\,b\,$, but only on
values of $\,\Delta_S\,$ and $\,y\,$. In the region of $\,b\,$,
from $\,b_{max}\,$ to the $\,b_0\,$ of \eq{L3},

$$
b_{0}\,>\,b\,>\,b_{max}\,\,,
$$
the amplitude is given by \eq{B9}.
There is a power-like decrease of the amplitude in this region
of $\,b\,$.
From the point $\,b\,=\,b_0\,$
and to the some value of $\,b\,$, which defines
the region of applicability of  the diffusion approach, and  which
is $\,b_{diff}\,=\,\,\sqrt{r_{1,t}\,r_{2,t}}\,e^{\,D\,y\,/2\,}\,$, 
the larger impact parameter behavior is governed  by usual 
BFKL amplitude \eq{L1}. Inserting the value 
$\,b_0\,$ of \eq{L3} in \eq{L1} we find, that the 
at this point the  intercept is very small:
$$
N_{BFKL}(y,r_{1, t},r_{2, t}; b)\,\sim\,
Const\,\frac{
e^{y\Le\,\omega_0\,-\,\frac{\Delta_S}{2^{8\,\nu_0\,}\,\nu_0}\,\Ra}}
{y^{1/4\nu_0}}\,. 
$$
For small values of $\,\nu_0\,$, the amplitude  
decreases with $\,y\,$. It is also possible, that
$\,b_0\,>\,b_{diff}\,$, and only the `effective'
Pomeron will define the large $\,b\,$ behavior.  
This picture
of the $\,b\,\,$ dependence is presented in the Fig.~\ref{REG}.

Returning to equation \eq{B9}
we see, that in the wide region of large $\,b\,$,
the `effective' Pomeron
preserves a  power-like decreasing structure and has 
the intercept $\,\omega_{S-H}\,$ larger than the 
intercept $\,\omega_0\,$ of the BFKL amplitude, 
see Section 5 of the paper.
Nevertheless, this decrease
is quite different from the structure of the  pure BFKL large $\,b\,$
asymptotic behavior. Indeed, in the  diffusion approach,
the BFKL amplitude leads to the $\,1/b^{2\,}\,$
behavior, see \eq{L1} while
the $\,b\,$ dependence of \eq{B9} is very different from this.
This  difference is in the sharper decrease, than the BFKL
amplitude has, but it is even more important, that 
the power of $\,b\,$ is governed by $\,\nu_0\,$ 
function which depends on the value of  $\,\Delta_S\,$,
i.e. by the soft, non-perturbative physics.
Only from very large $\,b\,$, defined by \eq{L3},
the asymptotic behavior of amplitude is again governed by
the BFKL amplitude. Indeed, it was also mentioned in Introduction, that
due the short range of the considered corrections,
the behavior of `effective ladder' from very large  $\,b\,$
will be again defined  by the BFKL amplitude. 
But still, the main  result of 
this consideration is that there is a region of
large $\,b\,$ where 
the admixture of the `soft' and `hard' Pomerons 
has the very different from the usual BFKL amplitude 
behavior.

 Another observation is that the results  of \eq{B6}
and \eq{B9} are obtained in the approximation
of small $\,\nu_0\,$, where $\,2\,\nu_0\,<\,1\,$
in the framework of the  diffusion approach for the BFKL kernel. 
It easy to see, that in the case when  $\,2\,\nu_0\,>\,1\,$
we will obtain a very different large $\,b\,$ behavior
from those of \eq{B6} or \eq{B9}, 
where the final answer will also   
depend on the value of $\,\nu_0\,$. 
In this case we need to include in the 
calculations the full BFKL kernel, instead of the diffusion approximation,
and this task is beyond  the framework of the paper.

{\large\bf Acknowledgments:}

I wish to thank E.Levin and E.Gotsman, without whose help
and advice this paper would not be written.

This research was supported by Israel Science Foundation, founded by
the Israeli Academy of Science and Humanities.  

\appendix
\section{}

  In this Appendix we consider the expression for the denominator
of \eq{SH1}, and will solve \eq{Pole}. In the
denominator of \eq{SH1} we must perform the integration over $\,k_1\,$, 
$\,k_2\,$, $\,r_{1,t}\,$ and $\,r_{2,t}\,$ variables. The integration 
over  $\,k_1\,$, $\,r_{1,t}\,$ is simple:

\beq\label{App1}
\int\,k_1^2\,d^2\,k_1\,\,
e^{\,-\frac{k^{2}_{1}}{q^{2}_{s}}+
\frac{\vec{k}_{1}\,\vec{q}}{q^{2}_{s}}}\,
\int\,\frac{d^2\,r_{1,t}}{(2\,\pi)^2}\,\,
e^{-i\,\vec{k}_1\,\vec{r}_1}\,\,
r_{1}^{1\,+\,2\,i\,\nu}\,=\,
\eeq 
$$
\,=\,\Le\,q_{s}\,\Ra^{1\,-\,2\,i\,\nu}\,2^{1\,+\,4\,i\,\nu}\,
\frac{\Gamma(\frac{3}{2}\,+\,i\,\nu)\,\Gamma(\frac{1}{2}\,-\,i\,\nu)}
{\Gamma(-\frac{1}{2}\,-\,i\,\nu)}\,
_{1}F_{1}(\frac{1}{2}\,-\,i\,\nu\,,\,1\,,\,\frac{q^2}{4\,q^{2}_{s}})\,.
$$

From the integrals over $\,k_2\,$, $\,r_{2,t}\,$, we see, that
the main contribution comes from  $\,r_{2,t}\,\sim\,1/q_s\,$. Assuming 
$\,q/q_s\,<\,1\,$ we replace                 
the $\,K_{2\,i\,\nu}\,$ function
by the first three terms of it's expansion over $\,r_{2,t}\,q\,$. 
We will see later, that three terms are enough in order to achieve the
desired precision $\,\Le\frac{q^2}{q_s^2}\Ra^{2\nu}\,$.
We have:

\beq\label{App2}
K_{2\,i\,\nu}(\frac{q\,r_{2,t}}{2})\,=
\eeq
$$
\,2^{-1\,+\,4\,i\nu}\,
\Le\,q\,r_{2,t}\,\Ra^{-2\,i\,\nu}\,\Gamma(2\,i\,\nu)\,+\,
2^{-1\,-\,4\,i\nu}\,
\Le\,q\,r_{2,t}\,\Ra^{2\,i\,\nu}\,\Gamma(-2\,i\,\nu)\,
+\,2^{-5\,+\,4\,i\nu}\,
\Le\,q\,r_{2,t}\,\Ra^{2-2\,i\,\nu}\,\frac{\Gamma(2\,i\,\nu)}
{1\,-\,2\,i\,\nu}\,.
$$

Inserting expression \eq{App2} back into the integral,
and integrating  over $\,k_2\,$, $\,r_{2,t}\,$, 
we obtain:

\beq\label{App3}
\int\,k_2^2\,d^2\,k_2\,
e^{\,-\frac{k^{2}_{1}}{q^{2}_{s}}+
\frac{\vec{k}_{1}\,\vec{q}}{q^{2}_{s}}}
\int\,\frac{d\,r_{2,t}}{(2\,\pi)}r_{2}^{2}J_{0}(k_1\,r_{1,t})
\eeq
$$
\Le
\,2^{-1\,+\,4\,i\nu}\,
\Le\,q\,r_{2,t}\Ra^{-2\,i\nu}\,\Gamma(2\,i\nu)\,+\,
2^{-1-4\,i\nu}
\Le\,q\,r_{2,t}\Ra^{2\,i\nu}\,\Gamma(-2\,i\nu)
\,+\,
\,2^{-5\,+\,4\,i\nu}\,
\Le\,q\,r_{2,t}\,\Ra^{2-2\,i\,\nu}\,\frac{\Gamma(2\,i\,\nu)}
{1\,-\,2\,i\,\nu}\,
\Ra\,=\,
$$ 
$$
\,=\,
2^{2\,i\,\nu}\,\frac{\Gamma(2\,i\,\nu)\,
\Gamma(3/2\,-\,i\,\nu)\Gamma(1/2\,+\,i\,\nu)}{\Gamma(-1/2\,+\,i\,\nu)}\,
q_{s}^{1\,+\,2\,i\,\nu}\,q^{-2\,i\,\nu}\,
_{1}F_{1}(\frac{1}{2}\,+\,i\,\nu\,,\,1\,,\,\frac{q^2}{4\,q^{2}_{s}})\,
+\,
$$
$$
\,+\,2^{-2\,i\,\nu}\,\frac{\Gamma(-2\,i\,\nu)\,
\Gamma(3/2\,+\,i\,\nu)\Gamma(1/2\,-\,i\,\nu)}{\Gamma(-1/2\,-\,i\,\nu)}\,
q_{s}^{1\,-\,2\,i\,\nu}\,q^{2\,i\,\nu}\,
_{1}F_{1}(\frac{1}{2}\,-\,i\,\nu\,,\,1\,,\,\frac{q^2}{4\,q^{2}_{s}})\,+
$$
$$
+\,
2^{-6\,+2\,i\,\nu}\,\frac{\Gamma(2\,i\,\nu)\,
\Gamma(5/2\,-\,i\,\nu)\Gamma(-1/2\,+\,i\,\nu)}
{\Le\,1\,-\,2\,i\,\nu\,\Ra\Gamma(-3/2\,+\,i\,\nu)}\,
q_{s}^{-1\,+\,2\,i\,\nu}\,q^{2\,-\,2\,i\,\nu}\,
_{1}F_{1}(\frac{1}{2}\,+\,i\,\nu\,,\,1\,,\,\frac{q^2}{4\,q^{2}_{s}})\,.
$$

So, now we are left  with the integral
over $\,\nu\,$:

\beq\label{App4}
\,\frac{\Delta_{S}}{\pi^2\,q_{s}^2}\,e^{-\frac{q^2}{q^{2}_{s}}}\,\,
\int\,\frac{d\,\nu\,q_{s}^{1\,-\,2\,i\,\nu}\,q^{2\,i\,\nu}\,
2^{5\,-\,4\,i\,\nu}}
{(\omega\,-\,\omega(\nu))\,(1/2\,-\,i\,\nu)}\,
\frac{\Gamma(\frac{3}{2}\,+\,i\,\nu)\,\Gamma(\frac{1}{2}\,-\,i\,\nu)}
{\Gamma(1\,+\,2\,i\,\nu)\,\Gamma(-\frac{1}{2}\,-\,i\,\nu)}\,
_{1}F_{1}(\frac{1}{2}\,-\,i\,\nu\,,\,1\,,\,\frac{q^2}{4\,q^{2}_{s}})\,
\eeq
$$
(\,
2^{2\,i\,\nu}\,\frac{\Gamma(2\,i\,\nu)\,
\Gamma(3/2\,-\,i\,\nu)\Gamma(1/2\,+\,i\,\nu)}{\Gamma(-1/2\,+\,i\,\nu)}\,
q_{s}^{1\,+\,2\,i\,\nu}\,q^{-2\,i\,\nu}\,
_{1}F_{1}(\frac{1}{2}\,+\,i\,\nu\,,\,1\,,\,\frac{q^2}{4\,q^{2}_{s}})\,
+\,
$$
$$
+\,
\,2^{-2\,i\,\nu}\,\frac{\Gamma(-2\,i\,\nu)\,
\Gamma(3/2\,+\,i\,\nu)\Gamma(1/2\,-\,i\,\nu)}{\Gamma(-1/2\,-\,i\,\nu)}\,
q_{s}^{1\,-\,2\,i\,\nu}\,q^{2\,i\,\nu}\,
_{1}F_{1}(\frac{1}{2}\,-\,i\,\nu\,,\,1\,,\,\frac{q^2}{4\,q^{2}_{s}})\,
\,+
$$
$$
+\,
2^{-6\,+2\,i\,\nu}\,\frac{\Gamma(2\,i\,\nu)\,
\Gamma(5/2\,-\,i\,\nu)\Gamma(-1/2\,+\,i\,\nu)}
{\Le\,1\,-\,2\,i\,\nu\,\Ra\Gamma(-3/2\,+\,i\,\nu)}\,
q_{s}^{-1\,+\,2\,i\,\nu}\,q^{2\,-\,2\,i\,\nu}\,
_{1}F_{1}(\frac{1}{2}\,+\,i\,\nu\,,\,1\,,\,\frac{q^2}{4\,q^{2}_{s}})\,
)\,.
$$

The terms with the gamma functions can be simplified.
For the first term in the bracket we have:

\beq\label{App5}
\frac{\Gamma(2\,i\,\nu)\,
\Gamma(\frac{3}{2}\,+\,i\,\nu)\,\Gamma(\frac{1}{2}\,-\,i\,\nu)
\Gamma(3/2\,-\,i\,\nu)\Gamma(1/2\,+\,i\,\nu)}
{\Gamma(1\,+\,2\,i\,\nu)\,\Gamma(-1/2\,-\,i\,\nu)\,
\Gamma(-1/2\,+\,i\,\nu)\,}\,
\,=\,
\frac{\pi\,(\,1/4\,+\,\nu^2)^{2}}{2\,i\,\nu\,\cosh(\pi\,\nu)}\,.
\eeq

According to \eq{App5},
the main contribution in the integral of \eq{App4} 
comes from the region of small $\,\nu\,$, therefore,
with $\,\omega(\nu)\,$ in diffusion approximation,
the first term in the bracket gives the following function
in the integral:

\beq\label{App6}
\frac{\pi\,(\,1/4\,+\,\nu^2)^{2}}{2\,i\,\nu\,\cosh(\pi\,\nu)\,
\Le\omega\,-\,\omega(\nu)\Ra\,\Le\,1/2\,-\,i\,\nu\,\Ra}\,
_{1}F^{2}_{1}(\frac{1}{2}\,+\,i\,\nu\,,\,1\,,\,\frac{q^2}{4\,q^{2}_{s}})\,
\approx\,
\frac{\pi}{16\,i\,\nu}\,
\frac{e^{\frac{q^2}{4\,q_{s}^{2}}}I_{0}^{2}(q^2/4q^{2}_{s})}{
\omega\,-\,\omega_{0}\,+\,D\,\nu^2\,}\,,
\eeq
where $\,I_{0}\,$ is the Bessel function of the second kind.
This answer is obtained in the  approximation
where $\,2\,\nu\,<\,1\,$. The case, when
$\,1<\,2\,\nu\,<\,2\,$, needs an additional consideration
and must include the full BFKL  kernel
in the calculations,  instead of the diffusion approximation.
This task is not in  the framework of this paper, and, therefore,
in the following only the case of small $\,\nu\,$
with  $\,2\,\nu\,<\,1\,$ will be considered.
In the limit of small $\,\nu\,$, the second term in the bracket yields
the same answer as \eq{App6}, with only change $\,i\rightarrow\,-i\,$,
and the third term gives:

\beq\label{App7}
\frac{9\,\pi}{(2^{11})\,i\,\nu}\,
\frac{e^{\frac{q^2}{4\,q_{s}^{2}}}I_{0}^{2}(q^2/4q^{2}_{s})}{
\omega\,-\,\omega_{0}\,+\,D\,\nu^2\,}\,.
\eeq

Now the equation \eq{Pole} has the form:

\beq\label{App8}
1\,-\,\frac{\Delta_{S}}{\pi\,}\,e^{-\,\frac{3\,q^2}{4\,q^{2}_{s}}}\,\,
\int\,\frac{d\,\nu\,2^{4\,-\,8\,i\,\nu}\,}
{\omega\,-\,\omega_{0}\,+\,D\,\nu^2\,}\,
\frac{I_{0}^{2}(q^2/4q^{2}_{s})}{\,i\,\nu}
\Le\,
\frac{1}{8\,}\,-\,\frac{1}{8\,}\,
\Le\,\frac{q^2}{q_s^2}\,\Ra^{2\,i\,\nu}\,+\,
\frac{9}{2^{11}\,}\,
\Le\,\frac{q^2}{q_s^2}\,\Ra\,\Ra\,=\,0\,.
\eeq

We close
the contour of integration in \eq{App8} in the lower semi-plane,
and after integration over $\,\nu\,$, we have the following equation
for the `effective' pole position :

\beq\label{App9}
1-\,\frac{2\,\Delta_S}{\,2^{\,8\,\nu_0}\,}
\frac{e^{-\frac{3q^2}{4\,q^{2}_{s}}}}{\omega-\omega_0\,}
+\,\frac{2\,\Delta_S}{\,2^{\,8\,\nu_0}\,}\,
\frac{e^{\frac{-3q^2}{4\,q^{2}_{s}}}}{\omega-\omega_0\,}\,
\Le\,\frac{q^2}{q_s^2}\,\Ra^{2\nu_0}+
\,\frac{9\Delta_S}{\,2^{7\,+\,8\,\nu_0}\,}\,
\frac{e^{-\frac{3q^2}{4\,q^{2}_{s}}}}{\omega-\omega_0\,}\,
\Le\frac{q^2}{q_s^2}\Ra\,=\,0\,,
\eeq
where $\,\nu_0\,=\,\sqrt{\frac{\omega\,-\,\omega_0}{D}}$.
The solution of this equation depends on the value of $\,\nu_0\,$.
There are two possible cases , which we consider separately.

\begin{enumerate}
\item The first solution we obtain assuming that $\,2\,\nu_0\,<\,1\,$
and $\,2\,\nu_0\,\ln(\frac{q_s^2}{q^2})\,<\,1\,$.
In this case the equation may be rewritten in the form:

\beq\label{App13}
1-\,\frac{2\,\Delta_S}{\,2^{\,8\,\nu_0}\,}
\frac{e^{-\frac{3q^2}{4\,q^{2}_{s}}}}{\omega-\omega_0\,}
\Le\,
1\,-\,e^{-\,2\,\nu_0\,\ln(\frac{q_s^2}{q^2})}
\,\Ra\,=\,0\,,
\eeq

which has the  following approximate  solution:

\beq\label{App14}
\omega\,=\,\omega_{S-H}\,=\,\omega_0\,+\,
\Le\,\Delta_S\,\Ra^{2}\,
\frac{\ln^{2}(\frac{q_s^2}{q^2})\,}{\,2^{\,16\,\nu_0\,-\,4}\,D}\,,
\eeq

with

\beq\label{App15}
\nu_0\,=\,\,\frac{4\,\Delta_S}{D}\,
\ln(\frac{q_s^2}{q^2})\,.
\eeq

\item The second solution is where
$\,2\,\nu_0\,<\,1\,$
and $\,2\,\nu_0\,\ln(\frac{q_s^2}{q^2})\,>\,1\,$.
Here we consider the  following terms in our main equation
\eq{App9}:

\beq\label{App16}
1-\,\frac{2\,\Delta_S}{\,2^{\,8\,\nu_0}\,}
\frac{e^{-\frac{3q^2}{4\,q^{2}_{s}}}}{\omega-\omega_0\,}
+\,\frac{2\,\Delta_S}{\,2^{\,8\,\nu_0}\,}\,
\frac{e^{\frac{-3q^2}{4\,q^{2}_{s}}}}{\omega-\omega_0\,}\,
\Le\,\frac{q^2}{q_s^2}\,\Ra^{2\nu_0}\,=\,0\,.
\eeq

The solution, which we obtain from \eq{App16}, is:

\beq\label{App17}
\omega\,=\,\omega_{S-H}\,=\,\omega_0\,+\, 
\,\frac{\Delta_S}{\,2^{\,8\,\nu_0\,-\,1}\,}\,-\,
\,2\Delta_S\,\Le\,\frac{q^2}{16\,q_s^2}\,\Ra^{2\nu_0}\,,
\eeq

where

\beq\label{App18}
\nu_0\,=\,
\sqrt{\,\frac{2\,\Delta_S}{D}\,}\,.
\eeq

\end{enumerate}

\section{}

The integral of \eq{B1} is

\beq\label{LB1}
N_{S-H}(y,r_{1, t},r_{2, t}; b)=
2\pi^2\,C\,r_{1,t}\,r_{2,t}\,\frac{\Delta_S}
{16^{4\,\nu_0\,-\,1}\,D}
(r_{1,t}\,q_s)^{2\,\nu_0}
\int^{q_s}_{0}\,q\,d\,q\,J_{0}(|q|\,|b|)\,
e^{\omega_{S-H}\,y}\,
\frac{K_{2\,\nu_0}(|q|\,|r_{2,t}|/2)}{\nu_0}
\Le\,\frac{q}{q_s}\Ra^{2\nu_0}\,.
\eeq

The value of this integral is defined by saddle
points of exponent, together with $\,J_0\,$ function, or
by initial point of integration. Initially, we consider
the saddle points of this integral, which depend
on the form of solution for $\,\Delta_{S-H}\,$,
and the $\,q\,$ region of applicability of this solution.

 Let us consider the solution for
$\,\omega_{S-H}\,$ given by \eq{HS2}. 
The contribution from this solution comes from  the region of $\,q\,$
defined by condition
$\,2\,\nu_0\,\ln(\frac{q_s^2}{q^2})\,<\,1\,$.
Taking the  
asymptotic expression for $\,J_{0}(q\,b)\,$ in \eq{LB1},
we have the saddle point equation:

\beq\label{LB3}
\frac{d\,\,}{d\,q\,\,}\Le\,\,\Delta_{S}^{2}\,\,y\,\,
\frac{\ln^{2}(\frac{q_s^2}{q^2})\,}{2^{16\nu_0\,-\,4}\,D}+\,i\,q\,b\,
\,\Ra\,=\,0\,,
\eeq
or

\beq\label{LB4}
\,-\,4\,\Delta_{S}^{2}\,\frac{\,y\,}{q}\,
\frac{\ln(\frac{q_s^2}{q^2})}{2^{16\nu_0\,-\,4}D}\,+\,i\,b\,\,=\,0\,.
\eeq
The solution of this equation can be obtained by
iteration and it is

\beq\label{LB5} 
q_{SP}^{0}\,=\,-4\,i\,
\,\Delta_{S}^{2}\,\frac{\,y\,}{2^{16\nu_0\,-\,4}\,b\,D}\,,\,\,
\eeq
\beq\label{LB6}
q_{SP}^{1}\,=\,-4\,i\,
\,\Delta_{S}^{2}\,\frac{\,y\,}{2^{16\nu_0\,-\,4}\,b\,D}\,
\ln\Le\,
\frac{b^2\,q_s^2\,D^2\,2^{16\nu_0\,-\,4}}
{16\,\Delta_{S}^{4}\,y^2\,}
\,\Ra\,.
\eeq
We see, that in this case for large $\,y\,$, we have $\,|q_{SP}\,b|\,
\propto\,y\,>>\,1\,$, therefore, the asymptotic expansion
of $\,J_0\,$ is, indeed, justified.
This solution gives:

\beq\label{LB8}
N_{S-H}(y,r_{1, t},r_{2, t}; b)\,\propto
\frac{r_{1,t}\,r_{2,t}^{1+2\nu_0}}{b^{2+2\nu_0}}\,
\,e^{\omega_0\,y\,+
y\,\frac{\Le\,\Delta_S\,\Ra^{2}}{2^{16\nu_0\,-\,4}\,D}\,
\ln^{2}\Le\,
\frac{2^{16\nu_0\,-\,4}\,q_s^2\,b^2\,D^2}
{\Le\,\Delta_S\,\Ra^{4}\,y^2\,}\,\Ra\,}\,\,.
\eeq
The condition of applicability of this solution,
$\,2\,\nu_0\,\ln(\frac{q_s^2}{q^2})\,<\,1\,$ requires
$\,|q_{SP}|\,>\,q_0\,=\,q_s\,e^{-\Le\frac{D}{2\,\Delta_S}\Ra^{1/2}}\,,$
therefore \eq{LB8} is applicable in the limited range of $\,b\,$:

\beq\label{LB7}
b\,<\,b_{max}\,=\,4\,
\,\Delta_{S}^{2}\,\frac{\,y\,}{q_s\,D\,2^{16\nu_0\,-\,4}}\,
\,e^{\Le\frac{D}{2\,\Delta_S}\Ra^{1/2}}\,,
\eeq
and reaches the maximum value at the point $\,b\,=\,b_{max}\,$: 

\beq\label{LB10}
N_{S-H}(y,r_{1, t},r_{2, t}; b)\,\propto\,
\frac{r_{1,t}\,r_{2,t}^{1+2\nu_0}}{b^{2+2\nu_0}_{max}}\,
\,e^{\omega_0\,y\,+\,y\frac{\Delta_S\,}{2^{16\nu_0\,-\,3}}}\,.
\eeq

For other $\,\omega_{S-H}\,$, given by \eq{HS3}, we have
from  \eq{LB1} the following
saddle point equation :

\beq\label{LB15}
\frac{d\,\,}{d\,q\,\,}\Le\,
-\,2\,\Delta_S\,y\,\,\Le\,\frac{q^2}{16\,q^2_s}\,\Ra^{2\,\nu_0}\,+\,
i\,q\,b\,\Ra\,=\,0\,,
\eeq
or
\beq\label{LB16}
-\,4\,\Delta_S\,y\,\,\frac{\,\nu_0\,}{q}
\Le\,\frac{q^2}{16\,q^2_s}\,\Ra^{2\,\nu_0}\,+\,
i\,b\,=\,0\,.
\eeq
The equation \eq{LB16} has the  approximate solution:

\beq\label{LB17}
q_{SP}\,=\,-\,i\,4\,q_s\,\Le\,
\frac{\,\Delta_S\,\nu_0\,y\,}{b\,q_s\,}
\,\Ra^{1\,+\,4\,\nu_0}\,,
\eeq
where again we  check that $\,|q_{SP}\,b\,|\,\propto\,y\,>\,1\,$.
This solution contributes in the region of $\,q\,$, where
$\,2\,\nu_0\,\ln(\frac{q_s^2}{q^2})\,<\,1\,$, this gives
$\,|q_{SP}|\,<\,q_0\,=\,q_s\,e^{-\Le\frac{D}{2\,\Delta_S}\Ra^{1/2}}\,$.
We obtain for \eq{LB1}:

\beq\label{LB18}
N_{S-H}(y,r_{1, t},r_{2, t}; b)\,\propto\,
\frac{1}{b^{2}\,q_s^2}\,
\,e^{\omega_0\,y\,+\,y\,\,\frac{\Delta_S}{2^{8\nu_0\,-\,1}}
-\,y\,2\,\Delta_S\,
\Le\,\frac{\,\Delta_S\,\nu_0\,y\,}{b\,q_s\,}\Ra^{4\nu_0}\,}\,.
\eeq

We now consider the contribution in our integral, from the region
of small $\,q\,$, which are close to the initial point of integration.
Initially, we suppose that $\,q\,$ is so small that $\,q\,b\,<\,1\,$.
Later, we will define the region of $\,b\,$ where this condition is 
satisfied. For such small $\,q\,$ we have, that 
$\,J_0(q\,b\,)\,\approx\,1\,$, and can expand $\,K_{2\,\nu_0}\,$
function around $\,q\,=\,0\,$.
We obtain:

\beq\label{LB19}
N_{S-H}(y,r_{1, t},r_{2, t}; b)=
2^{-1+4\nu_0}\pi^2\,r_{1,t}\,r_{2,t}\,
\frac{\Delta_S}{16^{4\,\nu_0\,-\,1}\,D\nu_0}
\Le\frac{r_{1,t}}{r_{2,t}}\Ra^{2\,\nu_0}\,
e^{\omega_{0}\,y\,+\,y\frac{\Delta_S}{8}}
\int^{\infty}_{0}\,d\,q^2\,\,e^{
-\,2\,\Delta_S\,y\,\,\Le\,\frac{q^2}{16\,q^2_s}\,\Ra^{2\,\nu_0}\,}\,\,.
\eeq
Here, in the region of small $\,q\,$, we used the solution given by 
\eq{HS3}, since our  function decreases strongly
with $\,q\,$, the integration over $\,q\,$ in \eq{LB19} 
goes  to  infinity.
We see from \eq{LB19}, that the main contribution in the  integral
comes from the region $\,q\,$ where
$\,q\,\sim\,4\,q_s\,\Le\,\frac{1}{2\,\Delta_S\,y}\,\Ra^{1/4\nu_0}\,$. 
Therefore, the condition $\,q\,b\,<\,1\,$ is satisfied for such $\,b\,$
which are
\beq\label{LB20}
4\,b\,q_s\,\Le\,\frac{1}{2\,\Delta_S\,y}\,\Ra^{1/4\nu_0}\,<\,1\,.
\eeq
In this region of $\,b\,$, after integration in \eq{LB19},
we obtain:

\beq\label{LB21}
N_{S-H}(y,r_{1, t},r_{2, t}; b)=
\frac{2^{4\nu_0}\pi^2}{16^{4\,\nu_0\,-\,1}}\,\sqrt{\frac{2\Delta_S}{D}}\,
\Le\,\frac{8}{\Delta_S\,y}\,\Ra^{1/2\nu_0}\,
\Le\,r_{1,t}\,r_{2,t}\,q_s^2\Ra\,
\Le\frac{r_{1,t}}{r_{2,t}}\Ra^{2\,\nu_0}\,
e^{\omega_{0}\,y+y\frac{\Delta_S}{2^{8\nu_0\,-\,1}}}\,
\Gamma(1+\frac{1}{2\nu_0})\,,
\eeq
where 
\beq\label{LB22}
\nu_0\,=\,\sqrt{\frac{2\,\Delta_S}{D}}\,.
\eeq
In the region of $\,b\,$, where

\beq\label{LB23}
4\,b\,q_s\,\Le\,\frac{1}{2\,\Delta_S\,y}\,\Ra^{1/4\nu_0}\,>\,1\,.
\eeq
in the integration
we can neglect the exponent and keep $\,J_0(qb)\,$ function. 
In this case the integration over $\,q\,$ is simple and gives:

\beq\label{LB24}
N_{S-H}(y,r_{1, t},r_{2, t}; b)=
\frac{2\,\pi^2}{16^{4\,\nu_0\,-\,1}}\,\sqrt{\frac{8\Delta_S}{D}}\,
\frac{\Le\,r_{1,t}\,r_{2,t}\,\Ra^{1+2\nu_0}\,}
{\Le\,b^2\,+\,r_{2,t}^2/4\,\Ra^{1+2\nu_0}\,}\,
e^{\omega_{0}\,y+y\frac{\Delta_S}{2^{8\nu_0\,-\,1}}}\,
\Gamma(1+2\nu_0)\,.
\eeq
Comparing \eq{LB21} and \eq{LB24} with \eq{LB8} and \eq{LB18},
and taking \eq{LB20}, \eq{LB23} into account, 
we see, that \eq{LB8} is suppressed at least
by factor $\,\frac{1}{b^4}\,$,
in comparison to  \eq{LB21}, and \eq{LB18}
is suppressed  at least by factor 
$\,\frac{e^{-\,y\,\,2\,\Delta_S\,
\Le\,\frac{\,\Delta_S\,\nu_0\,y\,}{b\,q_s\,}\Ra^{4\nu_0}\,}}
{y^{1/2\,\nu_0}}\,$, in comparison to  \eq{LB24}.
Therefore, in the following we take 
the expression obtained by the contribution  around the 
initial point of integration as the solution .

\section{}

The function 
$\,\tilde{N}^{\omega}(r_{1, t},r_{2, t}; q)\,$
is the Fourier transform of the function
$\,\tilde{N}^{\omega}(k_1,k_2; q)\,$.
Correspondingly to \cite{BON}
this function is defined as:

\beq\label{C1}
\tilde{N}^{\omega}(r_{1, t},r_{2, t}; q)\,=\,
\int\,d^2\,k_1\,\int\,d^2\,k_2\,
e^{i\,\vec{k}_1\,\vec{r}_{1, t}+i\,\vec{k}_2\,\vec{r}_{2, t}}\,
\,\tilde{N}^{\omega}(k_1,k_2; q)\,=\,
\eeq
$$
=\,\frac{\Delta_S}{q^2_s}\,
\int\,d^2\,k_1\,\int\,d^2\,k_2\,
e^{i\,\vec{k}_1\,\vec{r}_{1}+i\,\vec{k}_2\,\vec{r}_{2}}\,
\int\,d^2\,k'\,(k')^{2}\,\phi(k',q)\,
N_{0}^{\omega}(k',k_2; q)\,
\int\,d^2\,k''\,(k'')^{2}\,\phi(k'',q)\,
N_{0}^{\omega}(k_1,k''; q)\,,
$$
where, as in second section,
$\,N_{0}^{\omega}(k_1,k''; q)\,$ is the Fourier transform of

\beq\label{C2}
N_{0}^{\omega}(r_{1, t},r_{2, t}; q)\,
=\,
\,\Le\,\,r_{1,t}\,r_{2,t}\,\Ra\,
\int\,
\frac{d\,\nu\,}{2\,\pi}\frac{\,
\Le\,r^2_{1,t}\,q^2\,\Ra^{\,i\,\nu}\,
\,2^{4\,-\,8\,i\,\nu}\,}
{(\omega\,-\,\omega(\nu))\,\,(1/2\,-\,i\,\nu\,)}\,
\frac{K_{2\,i\,\nu}\,(|q|\,|r_{2,t}|/2)}{\Gamma(1\,+\,2\,i\,\nu)\,}\,\,,
\eeq
and the function $\,\phi\,$  is defined by \eq{SH2}.
From \eq{C1} we have

\beq\label{C3}
\tilde{N}^{\omega}(r_{1, t},r_{2, t}; q)\,=\,\Delta_S\,
\int\,d^2\,k_1\,k_1^{2}\,\phi(k_1,q)\,
N_{0}^{\omega}(k_1,r_{2, t}; q)\,
\int\,d^2\,k_2\,k_2^{2}\,\phi(k_2,q)\,
N_{0}^{\omega}(r_{1, t},k_2; q)\,=\,
\eeq
$$
=\frac{\Delta_S}{q^2_s}\,
\int\,d^2\,k_1\,k_1^{2}\phi(k_1,q)
\int\frac{d^2\,r^{'}}{(2\,\pi)^2}\,e^{-i\vec{k}_1\vec{r}^{'}}
N_{0}^{\omega}(r^{'},r_{2, t}; q)
\int\,d^2\,k_2\,k_2^{2}\phi(k_2,q)
\int\frac{d^2\,r^{''}}{(2\,\pi)^2}\,e^{-i\vec{k}_2\vec{r}^{''}}
N_{0}^{\omega}(r_{1,t},r^{''}; q)\,.
$$
Performing 
the same integration as in Appendix A
with the function \eq{C2},
we obtain with the precision of 
$\,\Le\,\frac{q}{q_s}\,\Ra^{2\,i\,\nu}\,\,$:

\beq\label{C4}
\tilde{N}^{\omega}(r_{1, t},r_{2, t}; q)\,=\,\Delta_S\,
\Le\,\,r_{1,t}\,r_{2,t}\Ra\,e^{-q^2/q_s^2}\,
\eeq
$$
\int\,
\frac{d\,\nu_1\,}{2\,\pi}\frac{
K_{2\,i\,\nu_1}\,(|q|\,|r_{2,t}|/2)\,
\Gamma(\frac{3}{2}\,+\,i\,\nu_1)\,\Gamma(\frac{1}{2}\,-\,i\,\nu_1)
\,2^{4\,-\,8\,i\,\nu_1}\,}
{(\omega\,-\,\omega(\nu_1))\,(1/2\,-\,i\,\nu_1\,)\,
\Gamma(1\,+\,2\,i\,\nu_1)\,\Gamma(-\frac{1}{2}\,-\,i\,\nu_1)}\,
\Le\,\frac{q}{q_s}\,\Ra^{2\,i\,\nu_1}\,
2^{1\,+\,4\,i\,\nu_1}\,
_{1}F_{1}(\frac{1}{2}\,-\,i\,\nu_1\,,\,1\,,\,\frac{q^2}{4\,q^{2}_{s}})\,
$$
$$
(\,\int\,
\frac{d\,\nu_2\,}{2\,\pi}\frac{\,
\Gamma(3/2\,-\,i\,\nu_2)\Gamma(1/2\,+\,i\,\nu_2)
\,2^{4\,-\,8\,i\,\nu_2}\,}
{(\omega\,-\,\omega(\nu_2))\,(1/2\,-\,i\,\nu_2\,)\,
(\,2\,i\,\nu_2)\,\Gamma(-1/2\,+\,i\,\nu_2)}\,
\Le\,r_{1,t}\,q_s\Ra^{2\,i\,\nu_2\,}\,
2^{2\,i\,\nu_2}\,
_{1}F_{1}(\frac{1}{2}\,+\,i\,\nu_2\,,\,1\,,\,\frac{q^2}{4\,q^{2}_{s}})\,+\,
$$
$$
\int
\frac{d\nu_2}{2\pi}\frac{
\Gamma(3/2+i\nu_2)\Gamma(1/2-i\nu_2)\,2^{4\,-\,8\,i\,\nu_2}\,}
{(\omega-\omega(\nu_2))\,(1/2-i\nu_2)
(-2i\nu_2)\Gamma(-1/2-i\nu_2)}
\Le\,r_{1,t}q\Ra^{2i\nu_2\,}
\Le\frac{q}{q_s}\Ra^{2i\nu_2}
2^{-2i\nu_2}\,
_{1}F_{1}(\frac{1}{2}-i\nu_2,1,\frac{q^2}{4\,q^{2}_{s}}))\,.
$$

\section{}

We have to calculate the following expression:

\beq\label{D1}
\tilde{N}^{\omega_{S-H}}(r_{1, t},r_{2, t}; q)\,=\,\Delta_S\,
\Le\,\,r_{1,t}\,r_{2,t}\Ra\,e^{-q^2/q^2_s}\,
\eeq
$$
\int\,
\frac{d\nu_1}{2\pi}\frac{
K_{2i\nu_1}\,(|q||r_{2,t}|/2)
\Gamma(\frac{3}{2}+i\nu_1)\Gamma(\frac{1}{2}-i\nu_1)
\,2^{4\,-\,8\,i\,\nu_1}\,}
{(\omega_{S-H}-\omega(\nu_1))\,(1/2-i\nu_1)
\Gamma(1+2i\nu_1)\Gamma(-\frac{1}{2}-i\nu_1)}
\Le\frac{q}{q_s}\Ra^{2\,i\,\nu_1}\,
2^{1\,+\,4\,i\,\nu_1}\,
_{1}F_{1}(\frac{1}{2}-i\nu_1\,,1,\,\frac{q^2}{4\,q^{2}_{s}})\,
$$
$$
(\int\,
\frac{d\nu_2}{2\pi}\frac{
\Gamma(3/2-i\nu_2)\Gamma(1/2+i\nu_2)\,2^{4\,-\,8\,i\,\nu_2}\,}
{(\omega_{S-H}-\omega(\nu_2))(1/2\,-\,i\,\nu_2)
(2i\nu_2)\Gamma(-1/2+i\nu_2)}
\Le\,r_{1,t}\,q_s\Ra^{2i\nu_2}
2^{2i\nu_2}\,
_{1}F_{1}(\frac{1}{2}+i\nu_2\,,\,1,\,\frac{q^2}{4\,q^{2}_{s}})+
$$
$$
\int
\frac{d\nu_2}{2\pi}\frac{
\Gamma(3/2+i\nu_2)\Gamma(1/2-i\nu_2)\,2^{4\,-\,8\,i\,\nu_2}\,}
{(\omega_{S-H}-\omega(\nu_2))\,(1/2-i\nu_2)
(-2i\nu_2)\Gamma(-1/2-i\nu_2)}
\Le\,r_{1,t}q\Ra^{2i\nu_2\,}
\Le\frac{q}{q_s}\Ra^{2i\nu_2}
2^{-2i\nu_2}\,
_{1}F_{1}(\frac{1}{2}-i\nu_2,1,\frac{q^2}{4\,q^{2}_{s}}))\,.
$$
where $\,\omega(\nu)\,=\,\omega_0\,-\,D\,\nu^2\,$.
We perform the integration over both variables $\,\nu_1\,$ and $\,\nu_2\,$
by closing the contours of integration in the lower half-plane,
and taking into account the  poles 
$\,\nu_1\,=\,-i\,\nu_0\,=-i\,\sqrt{\frac{\omega_{S-H}-\omega_0}{D}}\,$,
$\,\nu_2\,=\,-i\,\nu_0\,$ and $\,\nu_2\,=\,0\,$.
In the limit of small $\,\nu_0\,$
the integration over $\,\nu_1\,$
gives:

\beq\label{D2}
\frac{\,2^{5\,-\,8\,\nu_0}\,\,e^{\,q^2/8\,q_s^2}}{D\,\nu_0}\,
K_{2\,\nu_0}\,(|q|\,|r_{2,t}|/2)\,I_0(q^2/4q^2_s)\,
\frac{\Gamma(3/2)\,\Gamma(1/2)}{\Gamma(-1/2)}\,
\,\Le\,\frac{q}{q_s}\,\Ra^{2\,\nu_0}\,.
\eeq
Integration over $\,\nu_2\,$ gives:

\beq\label{D3}
\frac{\,2^{3\,-\,8\,\nu_0}\,e^{\,q^2/8\,q_s^2}}{D\,\nu_0^2}\,
I_0(q^2/4q^2_s)\,
\frac{\Gamma(3/2)\,\Gamma(1/2)}{\Gamma(-1/2)}\,
\Le\,r_{1,t}\,q_s\,\Ra^{2\,\nu_0}\,.
\eeq
Therefore, we obtain:

\beq\label{D4}
\tilde{N}^{\omega_{S-H}}(r_{1, t},r_{2, t}; q)\,=\,
\frac{\pi\,\Delta_S}{16^{4\,\nu_0\,-\,1}\,D\,\nu_0}\,
\frac{\Le\,\,r_{1,t}\,r_{2,t}\Ra}{\omega_{S-H}\,-\,\omega_0}\,
e^{-3q^2/4q^2_s}\,I_0^2(q^2/4q^2_s)\,
\Le\,\frac{q}{q_s}\,\Ra^{2\,\nu_0}\,
\Le\,r_{1,t}\,q_s\,\Ra^{2\,\nu_0}\,
K_{2\,\nu_0}\,(|q|\,|r_{2,t}|/2)\,.
\eeq

\end{document}